\documentclass[reprint, aip, footinbib,  letterpaper, superscriptaddress]{revtex4-2}

\usepackage[T1]{fontenc} 

\usepackage{amsmath,color, tikz}
\usetikzlibrary{matrix}
\usepackage{amssymb}
\usepackage{amsfonts}
\usepackage{amsthm}
\usepackage{mathtools}
\usepackage{comment}
\usepackage{hyperref}
\usepackage{cleveref}

\usepackage{pythonhighlight}

\usepackage{algorithm}
\usepackage{algpseudocode}
\usepackage{dsfont}

\usepackage{bbold}
\usepackage{chemformula}
\usepackage{siunitx}

\DeclarePairedDelimiterX\braket[2]{\langle}{\rangle}{#1 \delimsize\vert #2}
\DeclarePairedDelimiterX\braket3[3]{\langle}{\rangle}{#1 \delimsize\vert #2 \delimsize\vert #3}

\usepackage{accents}

\usepackage{fancyvrb}

\usepackage{pdfpages}

\newcommand{\vE}{\mathbf{E}}

\newcommand{\vB}{\mathbf{B}}

\newcommand{\vD}{\mathbf{D}}

\newcommand{\vnabla}{\boldsymbol{\nabla}}

\newcommand{\vH}{\mathbf{H}}

\newcommand{\vR}{\mathbf{r}}

\usepackage{listings}
\usepackage{xcolor}

\usepackage{makecell}       

\definecolor{codegreen}{rgb}{0,0.6,0}
\definecolor{codegray}{rgb}{0.5,0.5,0.5}
\definecolor{codepurple}{rgb}{0.58,0,0.82}
\definecolor{backcolour}{rgb}{0.95,0.95,0.92}

\lstdefinestyle{mystyle}{
    backgroundcolor=\color{backcolour},   
    commentstyle=\color{codegreen},
    keywordstyle=\color{magenta},
    numberstyle=\tiny\color{codegray},
    stringstyle=\color{codepurple},
    basicstyle=\ttfamily\footnotesize,
    breakatwhitespace=false,         
    breaklines=true,                 
    captionpos=b,                    
    keepspaces=true,                 
    showspaces=false,                
    showstringspaces=false,
    showtabs=false,                  
    tabsize=2
}

\makeatletter
\AtBeginDocument{\let\LS@rot\@undefined}
\makeatother

\begin{document}

	\title{FDTD with Auxiliary Bath Fields for Condensed-Phase Polaritonics: Fundamentals and Implementation}
	
	\author{Tao E. Li}%
	\email{taoeli@udel.edu}
	\affiliation{Department of Physics and Astronomy, University of Delaware, Newark, Delaware 19716, USA}

    \begin{abstract}
        Understanding condensed-phase polariton experiments requires accurately accounting for both realistic cavity geometries and the interplay between polaritons and material dark modes arising from microscopic molecular interactions. The finite-difference time-domain (FDTD) approach numerically propagates classical Maxwell's equations in the time domain, offering a versatile scheme for modeling polaritons in realistic cavities. However, the  simple dielectric functions routinely used in FDTD often fail to describe molecular details. Consequently, standard FDTD calculations, to date, cannot accurately describe processes involving the complex coupling between polaritons and dark modes, such as polariton relaxation,  transport, and condensation. For more faithful simulations of the energy flow between polaritons and dark modes, herein, local bath degrees of freedom coupled to the material polarization are explicitly included in FDTD to describe the dark-mode dynamics. This method --- FDTD with auxiliary bath fields (FDTD-Bath) --- is implemented in the open-source MEEP package by adding a Lorentz-Bath material susceptibility, where explicit bath modes are coupled to conventional Lorentz oscillators. With this Lorentz-Bath susceptibility,  linear polariton spectra and Rabi-splitting-dependent polariton relaxation rates in planar Fabry--P\'erot cavities are reproduced more accurately than those with the conventional Lorentz  susceptibility. Supported by a user-friendly Python interface and efficient MPI parallelism, the FDTD-Bath approach implemented in MEEP is ready to model a wide range of polariton phenomena involving realistic cavity geometries.
        \end{abstract}

	\maketitle

    \section{Introduction}

    For atomic or molecular transitions strongly coupled to cavity photon modes near resonance, the Rabi splitting observed in spectroscopic signals hallmarks the formation of hybrid light-matter states, known as polaritons \cite{Hopfield1958,Ebbesen2023,Hirai2023,Simpkins2023,Mandal2023ChemRev,Bhuyan2023,Ruggenthaler2023,Tibben2023,Xiang2024}. Under this strong coupling regime, experiments show that polariton formation provides a novel strategy to modify and engineer both light and matter properties --- ranging from creating quantum fluids of light \cite{Deng2010,Carusotto2013} to altering molecular energy transfer and reaction dynamics \cite{Hutchison2012,Thomas2016,Zhong2017,Thomas2019_science,Xiang2020Science,Chen2022,Ahn2023Science}. 

    To better model and understand polariton experiments, a variety of theoretical methods have been proposed over the years, including advanced quantum-optical models \cite{Galego2019,Campos-Gonzalez-Angulo2019,Hernandez2019,Hoffmann2020,Botzung2020,Mandal2020Polarized,LiHuo2021,DuDark2021,Fischer2021,YangCao2021,Wang2022JPCL,Poh2023,Suyabatmaz2023,Aroeira2023}, classical or semiclassical electrodynamics \cite{Sukharev2023a,Sukharev2023,Zhou2024}, multiscale molecular dynamics simulations \cite{Luk2017,Groenhof2019,Tichauer2021,Sokolovskii2022tmp, Li2020Water,Li2023QMMM,Li2024CavMD}, conventional or multicomponent electronic structure theory \cite{Flick2017,Haugland2020,Riso2022,Schafer2021,Bonini2021,Yang2021,Philbin2022,McTague2022,Liebenthal2022,Yu2024,Weight2024,Kuisma2022,Li2022JCTCNEO}, and exact quantum dynamics approaches \cite{Triana2020Shape,Rosenzweig2022,Gomez2023,Lindoy2024, SangiogoGil2024}. While each method  provides complementary insights toward understanding polariton experiments, significant challenges remain in establishing a unified computational framework for strong coupling: both complex multi-mode cavity geometries and a macroscopic number of realistic molecules should be treated on the equal footing \cite{Li2022Review, Fregoni2022,Ruggenthaler2023,Mandal2023ChemRev}.

    Among different theoretical schemes of polaritons, classical electromagnetism offers an attractive solution for modeling polariton experiments, because solving classical Maxwell's equations naturally accommodates complex cavity structures beyond the single-mode approximation widely applied in quantum optics. However, polariton experiments in the solid and liquid phases underscore the crucial role of molecular details in determining polariton dynamics, such as vibration-assisted exciton-polariton scattering \cite{Coles2011,Perez-Sanchez2025}, ballistic and diffusive polariton transport \cite{Balasubrahmaniyam2023,Xu2023Polariton}, and the role of IR-inactive modes under vibrational strong coupling \cite{Hirschmann2024,Ji2025}; not to mention polariton effects on chemical reactivity \cite{Hutchison2012,Thomas2016,Thomas2019_science,Ahn2023Science}. 
    Within classical electromagnetism, where molecular response is typically represented by dielectric functions of simple analytical forms, accurately capturing these polariton processes remains a persistent challenge.

    A key feature in condensed-phase polariton experiments is the presence of dark modes, asymmetric combinations of molecular motions which do not interact directly with external electromagnetic (EM) fields \cite{Li2022Review, Fregoni2022,Ruggenthaler2023,Mandal2023ChemRev,Simpkins2023,Xiang2024}. These dark modes nevertheless interact with the polariton states due to molecular interactions between the matter component of polaritons (known as the bright mode) and the dark modes themselves \cite{Du2018,Saez-Blazquez2018, Li2021Relaxation,Ji2025}. Due to these interactions, excited polariton states may transfer energy to the dark modes during polariton relaxation, whereas the excited dark modes can also infuse energy to the polariton states in the polariton transport and condensation processes \cite{Balasubrahmaniyam2023, Deng2010,Carusotto2013}.  Thus, accurately modeling the interplay between polaritons and dark modes is essential for simulating nonequilibrium polariton dynamics.

    To capture the interplay between polaritons and dark modes within classical electromagnetism, we extend the standard finite-difference time-domain (FDTD) algorithm \cite{Taflove2005} --- which solves Maxwell's equations on a Yee grid \cite{Yee1966} --- by explicitly including dark-mode degrees of freedom. In conventional FDTD, the material response is described by analytic dielectric functions, and the decay of material polarization is frequently modeled via a simple damping term \cite{Taflove2005}. In our FDTD with auxiliary bath fields (FDTD-Bath) approach, we instead couple the material polarization to a set of optically inactive bath oscillators for representing the dark modes. Unlike multiscale or cavity molecular dynamics simulations \cite{Luk2017,Groenhof2019,Tichauer2021,Sokolovskii2022tmp, Li2020Water,Li2023QMMM,Li2024CavMD}, the FDTD-Bath  approach does not directly simulate detailed molecular structures; instead, these microscopic molecular effects are incorporated phenomenologically via the bath-oscillator density of states and their coupling strengths to the material polarization.
    
    We  implement the FDTD-Bath approach in the open-source MIT Electromagnetic Equation Propagation (MEEP) package \cite{Oskooi2010} by introducing a Lorentz-Bath material susceptibility, where explicit bath degrees of freedom are coupled to conventional Lorentz oscillators in each grid point. With a hybrid C++/Python interface plus the MPI distributed-memory parallelism in MEEP, the FDTD-Bath approach remains both user friendly and computationally efficient.  Importantly, our approach enables direct tracking of  polariton and dark-mode dynamics in realistic cavity geometries, potentially providing a unique perspective for studying strong coupling phenomena such as polariton relaxation, transport, and condensation. 

    As the initial demonstration of the FDTD-Bath approach, this paper focuses on the underlying theory, implementation details, and a Python-based tutorial within MEEP. A few illustrative examples such as polariton relaxation dynamics in both one-dimensional (1D) and two-dimensional (2D) Fabry--P\'erot cavities are also presented. A comprehensive benchmark of this approach on studying various polariton processes will be reported separately.

    This paper is organized as follows. Sec. \ref{sec:theory} introduces the fundamental theory and working equations of the FDTD-Bath approach. Sec. \ref{sec:implementation} describes the implementation details in the MEEP package and the Python usage. Sec. \ref{sec:results} presents the simulation results.  We conclude in Sec. \ref{sec:conclusion}, and the Appendix summarizes the simulation parameters.

    \section{Theory}\label{sec:theory}

    We begin with the classical Maxwell's curl equations in a dielectric medium \cite{Griffiths1999}:
    \begin{subequations}
        \begin{align}
        \frac{\partial \vD(\vR, t)}{\partial t} &=  \vnabla\times \vH(\vR, t) ,  \\
        \frac{\partial \vB(\vR, t)}{\partial t} &= -\vnabla\times \vE(\vR, t) .
        \end{align}
    \end{subequations}
    The electric displacement field $\vD$ is related to the electric field $\vE$ by \begin{equation}
        \vD(\vR, t) = \epsilon_0\vE(\vR, t) + \mathbf{P}(\vR, t),
    \end{equation}
    where $\epsilon_0$ denotes the vacuum permittivity, and  $\mathbf{P}$ represents the polarization density of the medium. For non-magnetic materials, the magnetic  $\vH$ field is related to the magnetic  $\vB$ field by $\vH = \vB/\mu_0 $, with $\mu_0$ the vacuum permeability.

    Because the material polarization response generally depends on the field frequency, the polarization density $\mathbf{P}$ can be related to the electric field $\vE$ in the frequency domain by introducing the material susceptibility $\chi$:
    \begin{equation}\label{eq:PE_omega}
        \mathbf{P}(\vR,\omega) = \epsilon_0\chi(\vR, \omega) \mathbf{E}(\vR, \omega).
    \end{equation}
    The relative permittivity (dielectric function) of the material is then $\epsilon_{\rm r}(\vR, \omega) = 1 + \chi(\vR, \omega)$. In the time domain, Eq. \eqref{eq:PE_omega} becomes a causal convolution:
    \begin{equation}
        \mathbf{P}(\vR, t)  = \epsilon_0 \int_0^{t} \chi(\vR, t-\tau) \vE(\vR, \tau) d\tau .
    \end{equation}

    \subsection{The Lorentz oscillator model}

    In computational electrodynamics, the susceptibility $\chi(\vR, \omega)$ of realistic materials is usually represented  as a sum of simple analytical models --- most commonly Drude or Lorentz terms. To avoid complexity, here we focus on the Lorentz model for material susceptibility, which characterizes a  driven-damping harmonic oscillator. The equation of motion for the Lorentz polarization density reads:
    \begin{equation}\label{eq:EOM_Lorentz}
        \ddot{\mathbf{P}}(\vR, t) + \gamma \dot{\mathbf{P}}(\vR, t) + \omega_0^2\mathbf{P}(\vR, t) = \epsilon_0 \omega_{0}^2  \sigma(\vR) \vE(\vR, t),
    \end{equation}
    where $\omega_0$ denotes the oscillator frequency,  $\gamma$ represents the dissipation rate, and the dimensionless $\sigma(\vR)$ quantifies the local  coupling strength between the medium and the electric field at point  $\vR$.
    
    Fourier transforming Eq. \eqref{eq:EOM_Lorentz} gives
    \begin{equation}\label{eq:P_L_freq}
        -\omega^2 \mathbf{P}(\vR, \omega) - i\gamma\omega \mathbf{P}(\vR, \omega) + \omega_0^2 \mathbf{P}(\vR, \omega) = \epsilon_0 \omega_{0}^2 \sigma(\vR)  \vE(\vR, \omega).
    \end{equation}
    According to Eq. \eqref{eq:PE_omega}, reorganizing Eq. \eqref{eq:P_L_freq} yields the susceptibility of the Lorentz oscillator model: 
    \begin{equation}\label{eq:chi_L}
        \chi_{\rm{L}}(\vR, \omega) = \frac{  \omega_{0}^2 \sigma(\vR)}{ \omega_0^2 - \omega^2 - i\gamma\omega } .
    \end{equation}
    Setting $\omega_{0}^{2}=0$ in the denominator recovers the Drude model for free electrons.  In practice, one often fits experimental susceptibilities by combining multiple Lorentz and (or) Drude terms.

    \subsubsection{Impact on the polariton relaxation rate}
    When a dielectric medium forms polaritons with the cavity modes, the polariton relaxation rate, $\gamma_{\text{pol}}$,  inherits the dissipation rates from both the cavity mode ($\gamma_{\rm c}$) and the material ($\gamma_{\rm m}$) \cite{Deng2010}:
    \begin{equation}\label{eq:gamma_polariton}
        \gamma_{\text{pol}} = |X_{\rm c}|^2 \gamma_{\rm c} + |X_{\rm m}|^2 \gamma_{\rm m},
    \end{equation}
    where $|X_{\rm c}|^2$ and $|X_{\rm m}|^2$ denote the photonic and molecular weights in the polariton state, respectively, also known as the Hopfield coefficients \cite{Hopfield1958}. In the Lorentz medium, because the dissipation rate is a constant $\gamma_{\rm m} = \gamma$, at resonance conditions (when $|X_{\rm c}|^2 \approx |X_{\rm m}|^2 \approx 1/2$), the polariton relaxation rate $\gamma_{\text{pol}}$ becomes independent of  the Rabi splitting \cite{Vargas2024}. This behavior contrasts with recent multiscale and cavity molecular dynamics simulations \cite{Groenhof2019,Li2021Relaxation,Ji2025}, which explicitly include molecular dark-mode degrees of freedom and predict a Rabi-splitting-dependent polariton relaxation rate.

    \subsection{The Lorentz-Bath model}
    To achieve a more accurate description of polariton dynamics within classical electromagnetism, we use the following system-bath Hamiltonian density for light-matter interactions:
    \begin{subequations}\label{eq:H_full}
    \begin{equation}
        \mathcal{H} =   \frac{1}{2\rho} {\boldsymbol{\Pi}_\mathbf{P}}^2 + \frac{1}{2}\rho \omega_0^2\mathbf{P}^2  - \mathbf{P}\cdot{\mathbf{E}} + \mathcal{H}_{\text{EM}} + \mathcal{H}_{\rm{bath}}.
    \end{equation}
    Here, at each spatial point $\vR$, the polarization density $\mathbf{P}(\vR)$ acts as a three-dimensional harmonic oscillator of frequency $\omega_0$. The scalar density $\rho(\vR)$ is defined as $\rho(\vR) \equiv 1/(\epsilon_0 \sigma(\vR) \omega_0^2)$, where the dimensionless quantity $\sigma(\vR)$ has been introduced in Eq. \eqref{eq:EOM_Lorentz}. Interpreting $\mathbf{P}$ as a generalized coordinate, $\rho$ plays the role of its effective mass, and $\boldsymbol{\Pi}_\mathbf{P}$ is the corresponding canonical momentum.  The polarization density $\mathbf{P}$ is coupled to the local electric field $\vE$ via the linear relation $- \mathbf{P}\cdot{\mathbf{E}}$. Finally, $\mathcal{H}_{\text{EM}} = \frac{\epsilon_0}{2} \vE^2 + \frac{1}{2\mu_0} \vB^2$ represents the free-space EM Hamiltonian density.

    Since the field wavelength typically exceeds molecular dimensions, at each spatial point $\vR$ in EM simulations, the polarization density $\mathbf{P}$ represents a local collective molecular excitation coupled to the field, which can be referred to as the local bright mode. Consequently, at each spatial location $\vR$, other local molecular collective degrees of freedom that are decoupled from the external E-field can be modeled by a set of $N$ optically inactive bath oscillators $\{\mathbf{Y}_j\}$. The  Hamiltonian density for these bath oscillators is given by:
    \begin{equation}
    \begin{aligned}
        \mathcal{H}_{\rm{bath}} & =    \sum_{j=1}^{N} \left( \frac{1}{2\rho}{\boldsymbol{\Pi}_{\mathbf{Y}_j}}^2 + \frac{1}{2}\rho \omega_j^2\mathbf{Y}_j^2 - k_j  \mathbf{Y}_j \cdot \boldsymbol{\Pi}_{\mathbf{P}} \right) \\
        & + \frac{\rho}{2}\left(\sum_{j=1}^{N} k_j \mathbf{Y}_j \right)^2,
    \end{aligned}
    \end{equation}
    \end{subequations}
    where each bath oscillator indexed by $j$ has an intrinsic frequency $\omega_j$ and is coupled to the polarization density $\mathbf{P}$  with  strength $k_j$, and $\boldsymbol{\Pi}_{\mathbf{Y}_j}$ denotes the canonical momentum of each bath oscillator. These phenomenological couplings $\{k_j\}$ reflect underlying molecular interactions and can be parameterized from outside-cavity experiments or molecular dynamics simulations. According to Table \ref{table:bath_parameters} below, $\{k_j\}$ are determined by parameters such as bath density of states and the dephasing rate from the material polarization to the bath, all of which can be determined from either outside-cavity experiments or molecular dynamics simulations.
    
    Note that explicitly including a set of bath oscillators was previously introduced for a canonical quantization scheme of macroscopic electrodynamics in lossy and dispersive  dielectrics \cite{Huttner1992}. Simulating collective strong coupling using Hamiltonians analogously to Eq. \eqref{eq:H_full} under the single-mode limit has also been studied previously \cite{Li2024Supervibronic}.
    
    According to the Hamiltonian density in Eq. \eqref{eq:H_full}, the matter equations of motion become:
    \begin{widetext}
    \begin{subequations}\label{eq:EOM_LorentzBath}
        \begin{align}
            \ddot{\mathbf{P}}(\vR, t) + \gamma_0\dot{\mathbf{P}}(\vR, t) + \sum_{j} k_j \dot{\mathbf{Y}}_j (\vR, t) + \omega_0^2\mathbf{P}(\vR, t) & = \epsilon_0 \omega_{0}^2 \sigma(\vR)  \vE(\vR, t) ,  \label{eq:EOM_LorentzBath-1}\\
            \ddot{\mathbf{Y}}_j(\vR, t) + \gamma_{j} \dot{\mathbf{Y}}_{j}(\vR, t) + \omega_{j}^2\mathbf{Y}_{j}(\vR, t) & =  k_j \dot{\mathbf{P}}(\vR, t) . \label{eq:EOM_LorentzBath-2}
        \end{align}
    \end{subequations}
    \end{widetext}
    Here, a phenomenological damping term $\gamma_0\dot{\mathbf{P}}$ is added on  the polarization density dynamics [Eq. \eqref{eq:EOM_LorentzBath-1}] to account for the dissipation beyond the coupling to the bath oscillators. Similarly, in Eq. \eqref{eq:EOM_LorentzBath-2}, a phenomenological damping term $\gamma_{j} \dot{\mathbf{Y}}_{j}$ is also added on each bath oscillator to account for the dissipation to other degrees of freedom that are not explicitly accounted for in Eq. \eqref{eq:EOM_LorentzBath}. 
    
    Fourier transforming Eq. \eqref{eq:EOM_LorentzBath} yields
    \begin{widetext}
    \begin{subequations}\label{eq:EOM_LorentzBath_FT}
        \begin{align}
            -\omega^2\mathbf{P}(\vR, \omega) - i \gamma_0\omega\mathbf{P}(\vR, \omega) - \sum_{j} i k_j \omega \mathbf{Y}_j (\vR, t) + \omega_0^2\mathbf{P}(\vR, \omega) & = \epsilon_0 \omega_{0}^2 \sigma(\vR)  \vE(\vR, \omega) ,  \label{eq:FT_LorentzBath-1}\\
            -\omega^2\mathbf{Y}_j(\vR, \omega) - i \gamma_{j} \omega \mathbf{Y}_{j}(\vR, \omega) + \omega_{j}^2\mathbf{Y}_{j}(\vR, \omega)  & =  - i k_j \omega\mathbf{P}(\vR, \omega) . \label{eq:FT_LorentzBath-2}
        \end{align}
    \end{subequations}
    \end{widetext}
    Substituting Eq. \eqref{eq:FT_LorentzBath-2}, or $\mathbf{Y}_{j}(\omega) = \frac{-ik_j\omega }{\omega_{j}^2 - \omega^2 - i\gamma_{j}\omega  } \mathbf{P}(\omega)$, into Eq. \eqref{eq:FT_LorentzBath-1}, and also applying Eq. \eqref{eq:PE_omega},  we obtain the Lorentz-Bath susceptibility: 
    \begin{subequations}\label{eq:chi_LB}
        \begin{equation}
        \chi_{\rm{LB}}(\vR, \omega) =  \frac{ \omega_{0}^2 \sigma(\vR)}{ \omega_0^2 - \omega^2 - i\gamma_0 \omega -  \Sigma(\omega) } ,
    \end{equation}
    where the bath self-energy term $\Sigma(\omega)$ is defined as
    \begin{equation}\label{eq:chi_LB_self_energy}
        \Sigma(\omega) = \sum_{j}  \frac{k_j^2 \omega^2}{\omega_j^2 - \omega^2 - i\gamma_j\omega} .
    \end{equation}
    \end{subequations}
    
    By analogy with the pure Lorentz form [Eq. \eqref{eq:chi_L}], we may rewrite the the Lorentz-Bath susceptibility as
    \begin{equation}
        \chi_{\rm{LB}}(\vR, \omega) = \frac{ \omega_{0}^2 \sigma(\vR)}{ \widetilde{\omega}_0^2 - \omega^2 - i\gamma_{\rm{LB}}(\omega) \omega } ,
    \end{equation}
    where the frequency-dependent damping rate is
    \begin{equation}\label{eq:gamma_LB}
        \gamma_{\rm {LB}}(\omega) = \gamma_0 + \frac{1}{\omega}\text{Im}\left[ \Sigma(\omega) \right] ,
    \end{equation}
     and the renormalized oscillator frequency $\widetilde{\omega}_0$ is determined by
    \begin{equation}\label{eq:omega_LB}
        \widetilde{\omega}_0^2 = \omega_0^2 - \text{Re}\left[ \Sigma(\omega) \right] .
    \end{equation}

    \subsubsection{Evaluating the self-energy $\Sigma(\omega)$}
    When all bath oscillators share the same coupling strength and damping rate, i.e., when $k_j = k_{\rm b}$ and $\gamma_j = \gamma_{\rm b}$ for $j = 1,2,\cdots,N$, the self-energy term in Eq. \eqref{eq:chi_LB} can be approximated as an integral:
    \begin{equation}\label{eq:sum_to_int}
        \Sigma(\omega)
         \approx \int_{0}^{+\infty} dv \rho(v) \frac{k_{\rm b}^2 \omega^2}{v^2 - \omega^2 - i\gamma_{\rm b}\omega} ,
    \end{equation}
    where $\rho(v)$ denotes the density of states of the bath oscillators. If the density of states $\rho(v)$  varies slowly near $v \approx \omega$, one may take $\rho(v)\approx\rho(\omega)$ outside the integral. Then, using the contour integral identity  $\lim_{\gamma_{\rm b} \rightarrow 0}\int_{0}^{+\infty} dx \frac{ \omega^2}{x^2 - \omega^2 - i\gamma_{\rm b}\omega} =  i\frac{\pi\omega}{2}$, which is valid when $\gamma_{\rm b} \ll \omega$, we  obtain a simple analytical approximation of the self-energy:
    \begin{equation}\label{eq:int_countor_simplified}
                \Sigma(\omega) \approx \frac{i\pi }{2} k_{\rm b}^2  \omega\rho(\omega) .
    \end{equation}
     Substituting Eq. \eqref{eq:int_countor_simplified} into Eq. \eqref{eq:gamma_LB} yields a simple analytical form of the effective damping rate of the Lorentz-Bath model:
    \begin{equation}\label{eq:gamma_LB_simple}
        \gamma_{\rm{LB}}(\omega) \approx \gamma_0 + \frac{\pi}{2} k_{\rm b}^2 \rho(\omega) .
    \end{equation}
    Obviously, when the bath density of states obeys a uniform distribution, $\gamma_{\rm{LB}}(\omega)$ reduces to a constant, and the Lorentz-Bath model recovers the standard Lorentz form defined in Eq. \eqref{eq:chi_L}.

    \subsubsection{Impact on the polariton relaxation rate}
    When the bath oscillators model the dark modes in polariton dynamics, the corresponding density of states $\rho(\omega)$ is typically peaked around the bright-mode frequency $\omega_0$. Combining Eq. \eqref{eq:gamma_polariton} with Eq. \eqref{eq:gamma_LB_simple}, the polariton relaxation rate for the Lorentz-Bath model under strong coupling becomes
    \begin{equation}\label{eq:gamma_polariton_LB}
        \gamma_{\text{pol}}(\omega) = |X_{\rm c}|^2 \gamma_{\rm c} + |X_{\rm m}|^2 \gamma_{\rm 0} + |X_{\rm m}|^2 \frac{\pi}{2} k_{\rm b}^2 \rho(\omega).
    \end{equation}
    As the Rabi splitting increases, the polariton frequency $\omega$ moves away from the molecular absorption frequency $\omega_0$ outside the cavity, reducing the bath density of states $\rho(\omega)$ and thus suppressing the polariton decay rate $\gamma_{\text{pol}}$. This result is consistent with recent analytical studies of the Rabi-splitting-dependent polariton relaxation rate \cite{Li2021Relaxation,Chng2024}.

    A cartoon comparison between the Lorentz oscillator model and the Lorentz-Bath model is given in Fig. \ref{fig:demo}a.

    \subsection{FDTD working equations}
    The analysis above shows that the Lorentz-Bath susceptibility reproduces the desired Rabi-splitting-dependent polariton relaxation rates. Given the recovery of this fundamental mechanism, the Lorentz-Bath model may also be advantageous on simulating other polariton-related phenomena in the condensed phase. To leverage this potential in practical simulations, we now present the working equations for implementing the Lorentz-Bath model in FDTD. For clarity, the FDTD working equations for the Lorentz-Bath model are provided in the 1D form, and the extension to 2D and 3D follows analogously. 

    \begin{figure*}
	    \centering
	    \includegraphics[width=1.0\linewidth]{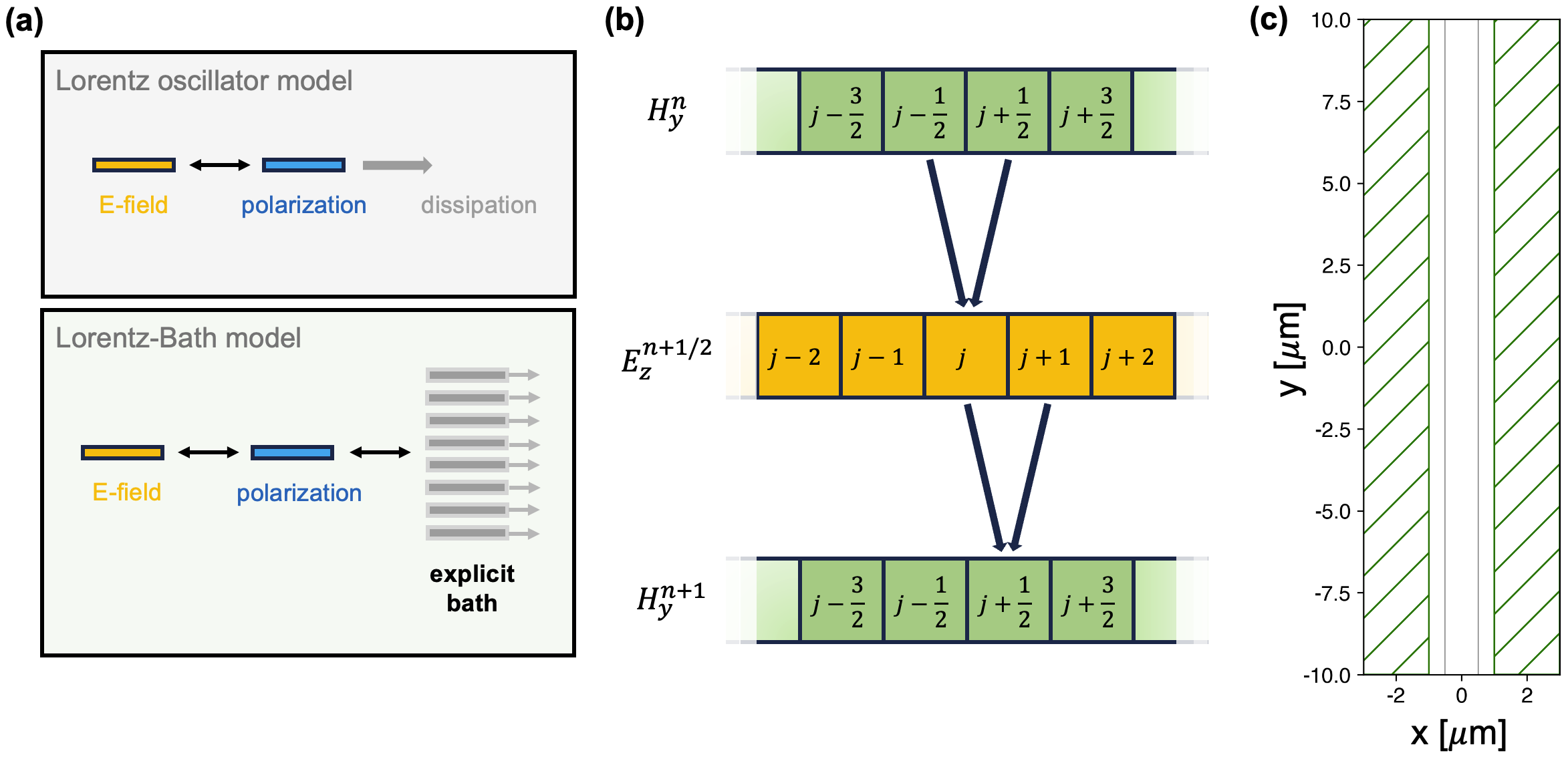}
	    \caption{(a) Comparison of the conventional Lorentz oscillator model and the Lorentz-Bath model: the latter explicitly incorporates  bath degrees of freedom coupled to the material polarization. (b) Schematic of the 1D FDTD algorithm employing the Yee cell in 1D.  (c) 2D simulation geometry for a planar Fabry--P\'erot cavity: Two thin dielectric mirrors (gray lines; each with thickness $0.02$ $\mu$m and refractive index $n=10$) are separated by $d=1$ $\mu$m. A uniform dielectric slab with thickness $d=1$ $\mu$m  modeled by either the Lorentz or the Lorentz-Bath susceptibility is placed in the cavity to form polaritons. The absorbing boundary conditions are applied along the $x$-axis using the perfectly matched layers (PML)  technique (dashed green areas), whereas the Bloch-periodic boundary conditions are applied along the $y$-axis. }
	    \label{fig:demo}
    \end{figure*}
    
    As sketched in Fig. \ref{fig:demo}b, the 1D FDTD loop updates fields on a grid of spatial step $\Delta x$ and time step $\Delta t$ \cite{Taflove2005}:
    \begin{widetext}
    \begin{subequations}\label{eq:FDTD-1D}
        \begin{align}
            D_z^{n+1/2}(j) &= D_z^{n-1/2}(j) + \frac{\Delta t}{\Delta x} \left[
            H_y^n\left(j+\frac{1}{2}\right) - H_y^n\left(j-\frac{1}{2}\right)
            \right] , \\
            E_z^{n+1/2}(j) &= \frac{1}{\epsilon_0} \left[
            D_z^{n+1/2}(j) - P_z^{n+1/2}(j) 
            \right ] , \\
            \label{eq:FDTD-1D-c}
            \text{Evaluate } & P_z^{n+3/2}(j) , \\
             H_y^{n+1}\left(j+\frac{1}{2}\right) &= H_y^{n}\left(j+\frac{1}{2}\right) - \frac{\Delta t}{\mu_0\Delta x} \left[
             E_z^{n+1/2}(j+1) - E_z^{n+1/2}(j)
             \right] .
        \end{align}
    \end{subequations}
    \end{widetext}
    Here, $n$ indexes the time steps  and $j$ indexes the spatial grid points. The electric or magnetic field is assumed to be polarized along the $z$- or $y$-direction, respectively, with the spatial grid points span the $x$-axis.
    For different dielectric media, the only difference lies in how the polarization $P_z$ is advanced [Eq. \eqref{eq:FDTD-1D-c}]. Note that boundary conditions and sources, which are implementation-specific, are omitted from this core FDTD loop. 

    \subsubsection{The Lorentz oscillator model}
    For the standard Lorentz model, the time-domain dynamics are governed by Eq. \eqref{eq:EOM_Lorentz}. Using the auxiliary differential equation (ADE) scheme \cite{Joseph1991}, we may use the following finite differences to approximate the time derivatives:
    \begin{subequations}\label{eq:finite-difference-scheme}
        \begin{align}
            \dot{\mathbf{P}}(n\Delta t) &\approx \frac{\mathbf{P}((n+1)\Delta t) - \mathbf{P}((n-1)\Delta t)}{2\Delta t} ,\\
            \ddot{\mathbf{P}}(n\Delta t) &\approx \frac{\mathbf{P}((n+1)\Delta t) - 2 \mathbf{P}(n\Delta t) + \mathbf{P}((n-1)\Delta t)}{\Delta t^2} .
        \end{align}
    \end{subequations}
    Substituting the above equations to Eq. \eqref{eq:EOM_Lorentz} and discretizing in 1D, we obtain the numerical scheme to update the value of $P_z^{n+1}$, $P_z$ at the $(n+1)$-th time step, using the values of $P_z$ in previous two time steps:
    \begin{equation}\label{eq:Pz_1D_Lorentz}
    \begin{aligned}
        P_z^{n+1}= & \frac{1}{1 + \frac{1}{2}\gamma\Delta t} \left[
        \epsilon_0 \omega_0^2\Delta t^2\sigma E_z^n - (\omega_0^2\Delta t^2 - 2)P_z^n \right. \\
        & \left .- (1 - \frac{1}{2}\gamma\Delta t) P_z^{n-1} 
        \right ] .
    \end{aligned}
    \end{equation}
    Here, $E_z^n$ denotes  $E_z$ at the $n$-th time step; the spatial dependence $(j)$ for $P_z$, $\sigma$, and $E_z$ is neglected for simplicity.
    When simulating the dielectric response of the Lorentz medium with  FDTD, we apply the numerical scheme in Eq. \eqref{eq:Pz_1D_Lorentz} to update the material polarization in Eq. \eqref{eq:FDTD-1D-c}. 

    \subsubsection{The Lorentz-Bath model}
    Extending the ADE approach to the Lorentz-Bath model, we first discretize the equation of motion for each bath oscillator.
    Applying the finite-difference scheme [Eq. \eqref{eq:finite-difference-scheme}] to $\dot{\mathbf{Y}}_j$, $\ddot{\mathbf{Y}}_j$, and $\dot{\mathbf{P}}$ in the dynamics of the bath oscillators [Eq. \eqref{eq:EOM_LorentzBath-2}], we obtain
    \begin{subequations}
        \begin{equation}\label{eq:Yzj_finite_difference}
        Y_{zj}^{n+1} = a_j Y_{zj}^n + (b_j + 1) Y_{zj}^{n-1} + c_j(P_z^{n+1} - P_z^{n-1}) ,
    \end{equation}
    where the corresponding parameters are defined as
        \begin{align}
            a_j &= \frac{2 - \omega_j^2\Delta t^2}{1 + \frac{1}{2} \gamma_j\Delta t} , \\
            b_j &= -\frac{2}{1 + \frac{1}{2} \gamma_j\Delta t} , \\
            c_j &= \frac{k_j\Delta t}{2\left(1 + \frac{1}{2} \gamma_j\Delta t \right)} .
        \end{align}
    \end{subequations}
    Next, to derive the numerical scheme for updating the polarization density [Eq. \eqref{eq:EOM_LorentzBath-1}], we apply the finite-difference formulas [Eq. \eqref{eq:finite-difference-scheme}] for $\dot{\mathbf{P}}$, $\ddot{\mathbf{P}}$, and $\dot{\mathbf{Y}}_j$. By further substituting $Y_{zj}^{n+1}$ [Eq. \eqref{eq:Yzj_finite_difference}] into Eq. \eqref{eq:EOM_LorentzBath-1}, we eventually find
    \begin{subequations}
        \begin{equation}\label{eq:Pz_update-LB}
        \begin{aligned}
            P_z^{n+1} = & \frac{1}{A}
         [ 
        \epsilon_0 \omega_0^2\Delta t^2\sigma E_z^n - (\omega_0^2\Delta t^2 - 2)P_z^n  \\
        &  - (1 -\frac{1}{2}\gamma_0 \Delta t - \frac{\Delta t}{2}\sum_j k_jc_j) P_z^{n-1} \\
        & 
        -\frac{\Delta t}{2} \sum_j k_j(a_j Y_{zj}^n + b_j Y_{zj}^{n-1})
         ] ,
        \end{aligned}
    \end{equation}
    with 
    \begin{equation}
        A = 1 + \frac{1}{2}\gamma_0 \Delta t + \frac{\Delta t}{2} \sum_{j} k_jc_j .
    \end{equation}
    \end{subequations}
    Overall, when simulating the dielectric response of the Lorentz-Bath model in FDTD, we use Eq. \eqref{eq:Pz_update-LB} at the polarization-update step [Eq. \eqref{eq:FDTD-1D-c}], while each bath oscillator is propagated using Eq. \eqref{eq:Yzj_finite_difference} during each time step.

    Although we have only presented the 1D implementation in detail, extending the Lorentz-Bath model to 3D  is  straightforward: For each polarization component ($P_x$, $P_y$, $P_z$), an independent set of 1D bath oscillators is coupled to that polarization component. All other aspects of the 3D FDTD algorithm, such as the updating scheme of the EM fields and the boundary conditions, follow the standard FDTD procedure.

    \section{Implementation Details}\label{sec:implementation}
    
    We integrated the Lorentz-Bath dielectric model into the open-source MEEP package \cite{Oskooi2010}. In the original MEEP package, the core FDTD engine was implemented in C++ and supported distributed-memory parallelism via MPI, with SWIG wrappers providing a high-level Python interface for setting up and running simulations.  Since MEEP already supported the Lorentz oscillator model through the C++ class \texttt{lorentzian\_susceptibility}, we implemented the Lorentz-Bath model by creating a derived C++ class, \texttt{bath\_lorentzian\_susceptibility}. On the Python side, the original MEEP package exposed the Lorentz functionality via the \texttt{LorentzianSusceptibility} class; we likewise added a derived Python class, \texttt{BathLorentzianSusceptibility}, so users can access the new Lorentz-Bath rountine directly from Python. 
    
    With this implementation strategy, switching from the standard Lorentz model to the Lorentz-Bath oscillator model in the Python API of MEEP requires only to change one line of code --- all other FDTD settings stay exactly the same. For example, in the MEEP Python interface, the conventional Lorentz susceptibility is normally defined as follows:
    \begin{lstlisting}[language=Python, label={lst:lorentz}, caption={MEEP code snippet for defining the conventional Lorentz susceptibility. The keywords \texttt{frequency}, \texttt{gamma}, and \texttt{sigma} correspond to the $\omega_0$, $\gamma$, and $\sigma$ parameters in the Lorentz susceptibility defined in Eq. \eqref{eq:chi_L}.}]
import meep as mp
mp.LorentzianSusceptibility(
        frequency=frequency_lorentz,
        gamma=gamma_lorentz,
        sigma=sigma_lorentz,
    )
\end{lstlisting}

\begin{table*}
  \caption{Connection between the bath parameters in
           Code Listing \ref{code:LB_lorentzian} and
           Code Listing \ref{code:LB_custom}. \footnote{Here, $\Delta \omega = \texttt{bath\_width}/(\texttt{num\_bath}-1)$ represents the frequency spacing between adjacent bath oscillators; $\Gamma = \texttt{gamma}\ (\gamma_0)  + \linebreak \texttt{bath\_dephasing}$ is set to the phenomenological dissipation rate of the material polarization ($\gamma_0$) plus the dephasing rate from the material polarization to the bath oscillators (\texttt{bath\_dephasing}).}}\label{table:bath_parameters}
  \centering
  \begin{tabular}{lcccr}
    \hline\hline
    \texttt{bath\_form} \ \  &
    $\{\omega_j\}$ (\texttt{bath\_frequencies}) &
    $\{\gamma_j\}$ (\texttt{bath\_gammas}) &
    $\{k_j\}$ (\texttt{bath\_couplings}) \\
    \hline
    \texttt{`uniform'} \ \  &
      \makecell[c]{%
        evenly spaced in \\
        $[\texttt{frequency} - \tfrac{\texttt{bath\_width}}{2},$ \\ $
          \texttt{frequency} + \tfrac{\texttt{bath\_width}}{2}]$} \ \  & \makecell[c]{[\texttt{bath\_gamma}, \texttt{bath\_gamma}, \\ $\cdots$, \texttt{bath\_gamma}]} \ \ 
      & $k_j \equiv k_{\rm b} = \sqrt{\frac{2\Delta \omega}{\pi} \times \texttt{bath\_dephasing}}$ \\[3pt]
    \texttt{`lorentzian'}  \ \ & same as above & same as above & $k_j= k_{\rm b} \sqrt{\frac{  \Gamma^2}{\Gamma^2 + (\omega_j - \omega_0)^2}}$ \\
    \texttt{`gaussian'}  \ \  & same as above & same as above & $k_j= k_{\rm b} \exp\left[-\frac{(\omega_j - \omega_0)^2}{4\Gamma^2}\right]$  \\
    \hline\hline
  \end{tabular}
\end{table*}

To switch to the Lorentz-Bath model, we simply replace the above Python class by:
\begin{lstlisting}[language=Python, caption={MEEP code snippet for defining the Lorentz-Bath susceptibility with customized bath oscillators. Similar to the Lorentz oscillator model, the keywords \texttt{frequency}, \texttt{gamma}, and \texttt{sigma} correspond to the $\omega_0$, $\gamma_0$, and $\sigma$ parameters in the Lorentz-Bath susceptibility defined in Eq. \eqref{eq:chi_LB}. The keyword \texttt{num\_bath} controls the number of bath oscillators included in the simulations. The Python lists \texttt{bath\_frequencies}, \texttt{bath\_gammas}, and \texttt{bath\_couplings} define the arrays $\{\omega_j\}$, $\{\gamma_j\}$, and $\{k_j\}$ for all the bath oscillators, respectively.}, label={code:LB_custom}]
mp.BathLorentzianSusceptibility(
        frequency=frequency_lorentz,
        gamma=gamma_lorentz*0.01,
        sigma=sigma_lorentz,
        num_bath=num_bath,
        bath_frequencies=[frequency_lorentz]*num_bath,
        bath_gammas=[gamma_lorentz*0.01]*num_bath,
        bath_couplings=[0.1]*num_bath,
    )
\end{lstlisting}

The above example requires the users to explicitly define the $\{\omega_j\}$, $\{\gamma_j\}$, and $\{k_j\}$ parameters for the bath oscillators, which can be tedious for beginners. Alternatively, the following simplified method for defining the Lorentz-Bath model is also available in MEEP:
\begin{lstlisting}[language=Python, caption={MEEP code snippet for defining the Lorentz-Bath susceptibility when the bath oscillators obey a uniform, Lorentzian, or Gaussian distribution. The parameters for the bath oscillators (with keywords starting from \texttt{bath\_}) are defined in Table \ref{table:bath_parameters}.  }, label={code:LB_lorentzian}]
mp.BathLorentzianSusceptibility(
        frequency=frequency_lorentz,
        gamma=gamma_lorentz*0.01,
        sigma=sigma_lorentz,
        num_bath=num_bath,
        bath_form="uniform", # or "lorentzian", "gaussian"
        bath_width=gamma_lorentz*10,
        bath_gamma=gamma_lorentz*0.01,
        bath_dephasing=gamma_lorentz*0.99,
    )
\end{lstlisting}

In this Python snippet, the variables \texttt{bath\_frequencies}, \texttt{bath\_gammas}, and \texttt{bath\_couplings} in Code Listing \ref{code:LB_custom} are omitted. When the keyword \texttt{bath\_form} is set to \texttt{"uniform"}, \texttt{"lorentzian"}, or \texttt{"gaussian"}, the bath oscillators obey the corresponding distribution, respectively. The detailed bath parameters are now controlled by the three new keywords, \texttt{bath\_width},   \texttt{bath\_gamma}, and \texttt{bath\_dephasing}, which define the bath oscillators as shown in Table \ref{table:bath_parameters}.

Among the three possible bath forms, the frequencies of the bath oscillators always obey a uniform distribution, evenly spaced across the interval [\texttt{frequency} - \texttt{bath\_width}/2, \texttt{frequency} + \texttt{bath\_width}/2]. Additionally, the dissipation rates of all the bath oscillators are assumed to be the same, which are controlled by the keyword \texttt{bath\_gamma}.

The keyword \texttt{bath\_dephasing} determines the dephasing (or energy transfer) rate from the material polarization to the bath oscillators in free space, given by $\frac{1}{\omega}\text{Im}[\Sigma(\omega)]$, where $\Sigma(\omega)$ denotes the self-energy term in Eq. \eqref{eq:chi_LB}. In free space, the overall linewidth of the Lorentz-Bath model should equal to $\Gamma = \texttt{gamma}\ (\gamma_0) + \texttt{bath\_dephasing} $,  the phenomenological dissipation rate of the material polarization ($\gamma_0$) plus the dephasing rate from the material polarization to the bath oscillators (\texttt{bath\_dephasing}).

Importantly, the bath distribution is controlled by the polarization-bath coupling $\{k_j\}$. For a uniform bath distribution,  Eq. \eqref{eq:int_countor_simplified} suggests that the keyword \texttt{bath\_dephasing} is related to the uniform polariton-bath coupling constant $k_{\rm b} = k_j$ ($\forall j$) via:
\begin{equation}\label{eq:bath_dephasing_value_uniform}
    \texttt{bath\_dephasing} = \frac{\pi}{2}k_{\rm b}^2\rho(\omega) = \frac{\pi}{2}k_{\rm b}^2 \frac{1}{\Delta \omega},
\end{equation}
where $\Delta \omega = \texttt{bath\_width}/(\texttt{num\_bath}-1)$ represents the frequency spacing between adjacent bath oscillators. Obviously, Eq. \eqref{eq:bath_dephasing_value_uniform} shows that assigning a value to \texttt{bath\_dephasing} effectively defines $k_{\rm b}$, or \texttt{bath\_couplings} in Code Listing \ref{code:LB_custom}. 

When the bath distribution becomes Lorentzian or Gaussian, as shown in Table \ref{table:bath_parameters}, $k_j^2$ is further modulated by a Lorentzian or Gaussian distribution with a characteristic linewidth $\Gamma$, matching that of the material polarization. It can be proved analytically that modulating  $k_j^2$  effectively defines a non-uniform bath density of states. 

For instance, substituting $k_j$ corresponding to the Lorentzian bath (Table \ref{table:bath_parameters}) into the self-energy in Eq. \eqref{eq:chi_LB_self_energy}, we obtain
\begin{subequations}
    \begin{align}
        \Sigma(\omega) &= \sum_{j} \frac{\Gamma^2}{\Gamma^2 +  (\omega_j - \omega_0)^2}  \frac{k_{\rm b}^2 \omega^2}{\omega_j^2 - \omega^2 - i\gamma_{\rm b}\omega} ,\\
        &\approx \int_{0}^{+\infty} dv \frac{1}{\Delta \omega}  \frac{\Gamma^2}{\Gamma^2 +  (v - \omega_0)^2}  \frac{k_{\rm b}^2 \omega^2}{v^2 - \omega^2 - i\gamma_{\rm b}\omega}  .
        \label{eq:self-energy-lorentz-bath-code}
    \end{align}
\end{subequations}
Here, we have replaced the sum by an integral using the uniform frequency density of states $1/\Delta \omega$. One then recognizes that Eq. \eqref{eq:self-energy-lorentz-bath-code} is  exactly the self-energy for a Lorentzian bath density of states $\rho(v) = \frac{1}{\Delta \omega}  \frac{\Gamma^2}{\Gamma^2 +  (v - \omega_0)^2}$ with constant bath decay rate $\gamma_j = \gamma_{\rm b}$  and uniform polarization-bath coupling $k_j=k_{\rm b}$ for all $j$.

As a consistency check, inserting the explicit form of $k_{\rm b}$  (Table \ref{table:bath_parameters})  in the integral above [Eq. \eqref{eq:self-energy-lorentz-bath-code}] and defining $I(\omega) = \int_{0}^{+\infty} dv \frac{\Gamma^2}{\Gamma^2 +  (v - \omega_0)^2}  \frac{\omega^2}{v^2 - \omega^2 - i\gamma_{\rm b}\omega}$, we find
\begin{equation}
        \Sigma(\omega) 
        = \frac{2}{\pi}I(\omega) \times \texttt{bath\_dephasing}.
\end{equation}
In the regime  $\Gamma, \gamma_{\rm b} \ll \omega_0$, the imaginary part of $I(\omega)$ admits the approximation $\text{Im}[I(\omega)] \approx \frac{\pi \omega}{2} \frac{\Gamma(\Gamma + \gamma_{\rm b}/2)}{(\omega-\omega_0)^2 + (\Gamma + \gamma_{\rm b}/2)^2}$. Setting the oscillator frequency $\omega = \omega_0$, i.e., in the absence of the polariton formation, $\text{Im}[I(\omega)] \approx \frac{\pi \omega}{2} \frac{\Gamma}{ \Gamma + \gamma_{\rm b}/2}$, so that $\frac{1}{\omega}\text{Im}[\Sigma(\omega)] \approx \texttt{bath\_dephasing}$ when $\Gamma \gg \gamma_{\rm b}$. Note that setting $\Gamma \gg \gamma_{\rm b}$ is essential for capturing the inverse energy transfer from the bath oscillators to the polaritons.  This proof confirms that our choice of parameters reproduces the desired overall polarization-bath dephasing rate.

    \section{Results}\label{sec:results}

    The FDTD-Bath approach introduced in this manuscript enables the direct simulation of polariton spectroscopy and dynamics in the basis of bright and dark modes across a wide variety of photonic environments. Below we present a few representative examples to illustrate the capacities of the FDTD-Bath approach; detailed simulation parameters are given in the Appendix.

    \subsection{Free-space spectra}

    Fig. \ref{fig:1d_results}a plots the 1D FDTD transmission spectra of a dielectric slab of thickness $d=1$ $\mu$m. When the dielectric slab is modeled by the conventional Lorentz susceptibility (gray), the  spectrum  peaks at $\omega_0 / 2\pi = 1$ $\mu$m$^{-1}$  with a linewidth  $\gamma /2\pi  = 4\times 10^{-2}$ $\mu$m$^{-1}$, in agreement with the parameters listed in the Appendix \footnote{Throughout this manuscript, all frequencies  are reported as $\nu = \omega/2\pi=\cdots$ because MEEP  defines the frequencies in terms of the temporal frequency $\nu$ instead of the angular frequency $\omega$.}. When the Lorentz-Bath model is applied with a uniform bath distribution (dashed cyan), the corresponding spectrum becomes visually identical to that of the Lorentz oscillator model. However, imposing a Lorentzian bath distribution centered at $\omega_0 /2\pi = 1$ $\mu$m$^{-1}$ (solid red) produces a more Gaussian-like lineshape, with suppressed intensity at the two tails of the lineshape.

    It is well known that homogeneous broadening yields Lorentzian profiles, while inhomogeneous broadening produces Gaussian-like spectra \cite{Knapp1981}.  Obviously, by tuning the distribution of the bath oscillators, the Lorentz-Bath susceptibility seamlessly interpolates between these two limits. Since the Lorentz-Bath model with a uniform bath reproduces the Lorentz oscillator model, we henceforth compare only the standard Lorentz oscillator model and the Lorentz-Bath model with a Lorentzian bath distribution, denoted as the Lorentz-Bath(L) model for brevity. Note that the results for the Lorentz-Bath model with a Gaussian bath distribution are not discussed here for simplicity, as this model yields very similar results as those of the Lorentz-Bath(L) model across this manuscript.

    \subsection{Collective strong coupling in a 1D Fabry--P\'erot cavity}

    \begin{figure}
	    \centering
	    \includegraphics[width=1.0\linewidth]{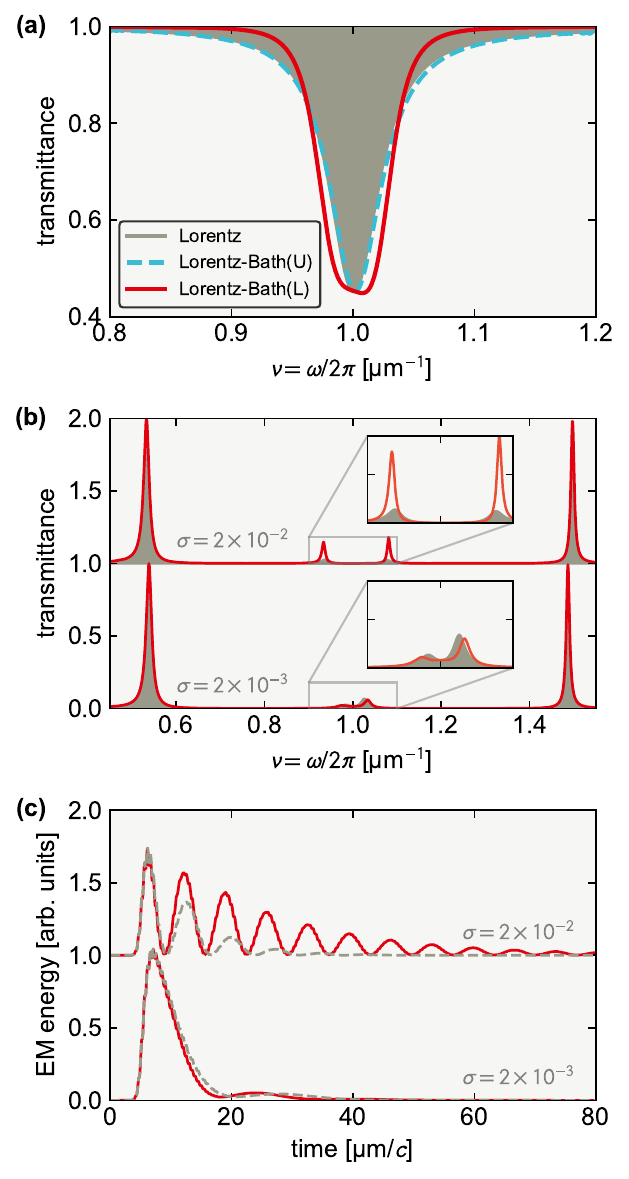}
	    \caption{Simulation results in 1D. (a) Free-space transmission spectra of a material slab modeled with different dielectric models: (i) the Lorentz oscillator model (gray shading), and the Lorentz-Bath model with bath modes following a (ii) uniform distribution (dashed cyan) or  (iii) Lorentzian distribution (solid red). (b) Cavity transmission spectra of the same dielectric slab confined in a Fabry--P\'erot cavity. The oscillator strength $\sigma$ of the dielectric layer is tuned to either $\sigma = 2\times 10^{-3}$ (bottom) or $2\times 10^{-2}$ (top). (c) Time-dependent EM energy dynamics inside the cavity under excitation of a broad-band Gaussian pulse. Compared to the standard Lorentz susceptibility, the Lorentz-Bath model yields stronger polariton transmission (panel b) and longer polariton lifetimes (panel c)  under large Rabi splittings.    }
	    \label{fig:1d_results}
    \end{figure}

    We consider polariton formation when the dielectric slab of width $d=1$ $\mu$m  is confined within the 2D planar Fabry--P\'erot cavity  illustrated in Fig. \ref{fig:demo}c. Owing to translational invariance along the $y$-axis, this dimension can be reduced, and 1D FDTD simulations can be applied for efficient computation. 

    Fig. \ref{fig:1d_results}b compares the linear transmission spectra of the strong coupling system using 1D FDTD simulations. When the effective coupling strength is set to $\sigma = 2\times 10^{-3}$ (bottom panel), the Lorentz-Bath(L) model (red) and  the conventional Lorentz model (gray) yield very similar results. Both spectra contain two uncoupled cavity modes at $\omega/2\pi = 0.5$ and $1.5$ $\mu$m$^{-1}$, respectively, while a pair of lower and upper polariton (LP and UP) peaks arises from resonance strong coupling between the cavity mode at $\omega/2\pi = 1.0$ $\mu$m$^{-1}$ and the confined dielectric slab. Remarkably, when the effective coupling strength is increased to $\sigma = 2\times 10^{-2}$ (top panel), the Lorentz-Bath(L) model predicts significantly enhanced polariton transmission peaks with substantially narrower linewidths than those of the Lorentz model, underscoring the role of the bath distribution in shaping polariton spectra.
    
    The enhanced polariton transmission predicted by the Lorentz-Bath(L) model can be understood as follows: At large Rabi splittings, the polariton frequencies lie in regions of reduced bath density of states, so the material polarization is only weakly coupled to the dark-mode oscillators.  Consequently,  polariton dissipation into the bath is suppressed, yielding narrower linewidths and stronger transmission --- much like the uncoupled cavity modes.

    We further understand the polariton signals by tracking the time-domain EM energy dynamics inside the cavity when the strong coupling system is excited by a broad-band Gaussian pulse. Because this Gaussian pulse  is centered at $\omega / 2\pi = 1$ $\mu$m$^{-1}$, both the UP and LP can be equally excited. As shown in the bottom part of Fig. \ref{fig:1d_results}c, when the effective coupling strength is $\sigma = 2\times 10^{-3}$, both the conventional Lorentz and Lorentz-Bath(L) models exhibit similar Rabi oscillations, confirming equal excitation of the UP and LP. However, when $\sigma$ is increased to $2\times 10^{-2}$, the Lorentz-Bath(L) model predicts a markedly longer-lived Rabi oscillation pattern compared to that of the Lorentz model. This longer oscillation period corresponds directly to increased polariton lifetimes and hence narrower polariton linewidths in the spectra, as shown in Fig. \ref{fig:1d_results}b.

    These numerical results agree with our analytical expression in Eq. \eqref{eq:gamma_polariton_LB}: For a Lorentzian bath distribution  centered at $\omega_0 / 2\pi= 1$ $\mu$m$^{-1}$, the Lorentz-Bath(L) model predicts reduced polariton decay rates at large Rabi splittings. This behavior is also consistent with prior atomistic polariton simulations in which the dark modes are modeled by real molecules \cite{Groenhof2019,Li2021Relaxation}.

    \subsection{Collective strong coupling in a 2D Fabry--P\'erot cavity}

    \begin{figure}
	    \centering
	    \includegraphics[width=1.0\linewidth]{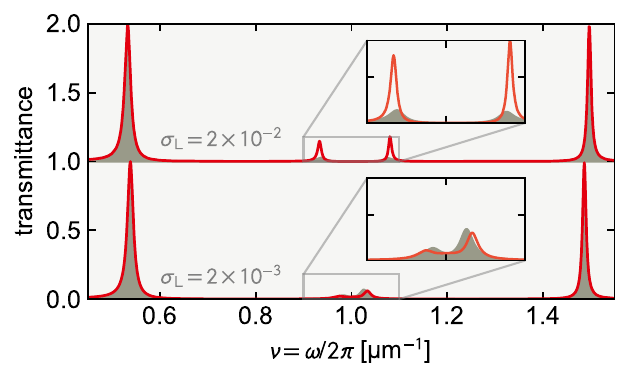}
	    \caption{Cavity transmission spectra of a dielectric slab confined  in a 2D Fabry--P\'erot cavity computed using 2D FDTD. The dielectric slab is modeled with either the Lorentz oscillator model (gray) or the Lorentz-Bath model with a Lorentzian bath distribution (red).  These 2D FDTD spectra are visually indistinguishable from the 1D FDTD results shown in Fig. \ref{fig:1d_results}b, confirming  the robustness of the FDTD-Bath approach across simulations in different dimensions. }
	    \label{fig:2d_results}
    \end{figure}
    
    To further verify the robustness of our FDTD-Bath implementation, we also perform full 2D FDTD simulations of the strong coupling system shown in Fig. \ref{fig:demo}c. As plotted in Fig. \ref{fig:2d_results}, the 2D transmission spectra at normal incidence are visually identical to those in 1D (Fig. \ref{fig:1d_results}b). This dimensional consistency confirms that our implementation yields valid and reliable predictions across different simulation geometries.

    \subsection{Parallel performance}

    \begin{figure}
	    \centering
	    \includegraphics[width=1.0\linewidth]{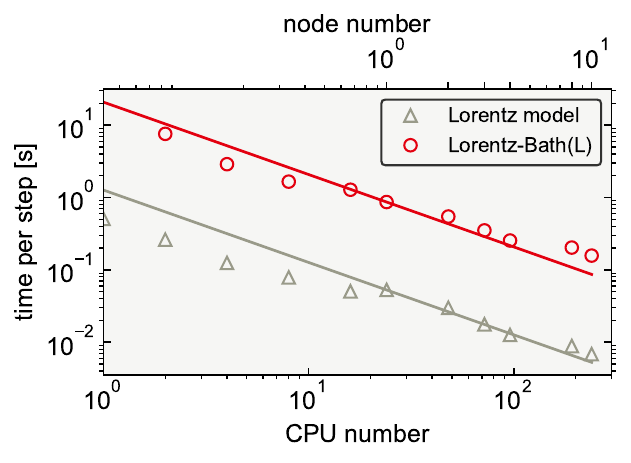}
	    \caption{Computational runtime per FDTD step as a function of the CPU core count for 2D polariton calculations. The confined dielectric slab is modeled by the Lorentz oscillator model (gray triangles) or the Lorentz-Bath model with a Lorentzian bath distribution (red circles). Solid lines denote the ideal inverse-linear parallel scaling. Benchmarks employ up to ten computational nodes (each with 24 CPU cores) to simulate the 2D cavity in Fig. \ref{fig:demo}c on a grid of $N_{\rm{grid}}=1.92\times 10^7$  points. As the Lorentz-Bath model includes $10^2$ bath oscillators in each grid point, the corresponding computational cost is approximately 20 times more expensive than that of the Lorentz oscillator model. 
  }
	    \label{fig:mpi_benchmark}
    \end{figure}

        While Fig. \ref{fig:2d_results} demonstrates compatibility of the FDTD-Bath approach with both 1D and 2D simulations,  realistic  polariton modeling also demands an efficient parallel computation scheme. Fig. \ref{fig:mpi_benchmark} reports the MPI parallel performance for the 2D strong coupling system depicted in Fig. \ref{fig:demo}c using our modified MEEP code. With a total grid size of $N_{\rm{grid}}=1.92\times 10^7$ spatial grid points in FDTD calculations, the simulations were run on up to ten nodes of the Caviness High-Performance Computing cluster at the University of Delaware, where 24 CPU cores were used in each node.   Overall, 
        both the Lorentz-Bath(L) model (red circles) and the standard Lorentz model (gray triangles) exhibit near-ideal inverse scaling with CPU count up to 240 cores, indicating that adding the bath fields does not degrade parallel efficiency. 
        
        In absolute terms, the Lorentz-Bath susceptibility incurs approximately a 20 times increase in runtime compared to that of the Lorentz model. This runtime increase is not surprising, since the Lorentz-Bath susceptibility embeds $10^2$ bath oscillators at each grid point. With a total grid size of $N_{\rm{grid}}=1.92\times 10^7$, this calculation involves approximately $10^9$ bath oscillators (or material dark modes). As a side note, using $10^2$  oscillators per grid point may be redundant: comparable accuracy can be obtained with far fewer bath modes by appropriately optimizing the bath parameters. The detailed parameter optimization in the FDTD-Bath approach will be reported separately.

    \section{Conclusion}\label{sec:conclusion}

    To summarize, we have developed and implemented the FDTD-Bath approach for simulating the interplay between polaritons and material dark  modes in realistic photonic environments. By explicitly including bath degrees of freedom coupled to the material polarization, our approach has accurately reproduced linear polariton spectra and Rabi-splitting-dependent polariton decay rates in agreement with prior studies.  We have also highlighted the open-source MEEP implementation of our approach, especially its intuitive Python interface and scalable MPI parallelism. To make practical use of the FDTD-Bath approach, the phenomenological parameters governing the bath fields must be obtained from outside-cavity experiments or molecular simulations. Once these parameters are determined, however, the FDTD-Bath framework can be readily applied to model polariton phenomena across a wide variety of realistic cavity geometries, spanning from the experimentally observed motional narrowing of  exciton-polariton linewidths \cite{Wanasinghe2024,Odewale2024} to more exotic nonequilibrium polariton dynamics.

    This manuscript presents the first realization of the FDTD-Bath approach using the Lorentz-Bath susceptibility.  While this work focuses only on a harmonic bath, including anharmonicity and thermal effects in the bath oscillators may 
    bring us a step closer to the elusive goal of faithfully capturing the interplay between polaritons and  dark modes in realistic, experimentally relevant conditions. Building on this foundation, layering explicit bath fields onto other widely applied susceptibilities, such as electronic two-level systems, may offer a computationally efficient strategy to study exciton-polariton dynamics in the condensed phase. Looking forward, we aim to explore how the FDTD-Bath approach can advance our understanding of complex polariton experiments.

    \section{Acknowledgments}
    This material is based upon work supported by the U.S. National Science Foundation under Grant No. CHE-2502758. This research is supported in part through the use of Information Technologies (IT) resources at the University of Delaware, specifically the high-performance computing resources.  This work also used the Anvil HPC at Purdue University through allocation CHE250091 from the Advanced Cyberinfrastructure Coordination Ecosystem: Services \& Support (ACCESS) program, which is supported by U.S. National Science Foundation grants \#2138259, \#2138286, \#2138307, \#2137603, and \#2138296.

    \section{Data Availability Statement}
    The modified MEEP code for the FDTD-Bath approach is available at the following Github repository: \url{https://github.com/TaoELi/meep}. The input and post-processing files of this manuscript are archived in a separated Github repository: \url{https://github.com/TaoELi/fdtd_bath}.
    
    \appendix*

    \setcounter{table}{0}
	\setcounter{equation}{0}
	\renewcommand{\theequation}{S\arabic{equation}}

	\maketitle
	
    
    \section{Simulation details}

    In the 1D FDTD simulations, the length of the simulation cell was 6 $\mu$m with spatial resolution of $\Delta x = 10^{-2}$ $\mu$m. The perfectly matched-layer (PML) absorbing boundary conditions, each 0.5 $\mu$m thick, were applied at both ends of the simulation cell. The transmission spectra were obtained by evaluating the flux spectrum in response to a broad-band Gaussian pulse centered at $\omega/2\pi = 1$ $\mu$m$^{-1}$ with a spectral width of 1.2 $\mu$m$^{-1}$. The external Gaussian pulse was applied at the left end of the simulation cell to excite the material system. For the free-space spectra calculations, a dielectric slab of thickness $d=1$ $\mu$m was placed at the center of the simulation cell. The dielectric medium was modeled by either the Lorentz or the Lorentz-Bath susceptibility. 
    
    For the Lorentz susceptibility, the following set of parameters were used for the free-space spectra: $\omega_0/2\pi = 1$ $\mu$m$^{-1}$, $\gamma/2\pi = 4\times 10^{-2}$ $\mu$m$^{-1}$, and $\sigma = 5\times 10^{-3}$. In the Lorentz-Bath calculations, the phenomenological relaxation rate of the material polarization was set to $\gamma_0/2\pi = 4\times 10^{-4}$, and the dephasing rate from the material polarization to the bath oscillators was set to $\gamma_{\rm{dephasing}}/2\pi = (\gamma - \gamma_0)/2\pi = 3.96\times 10^{-4}$ $\mu$m$^{-1}$, so that the total transmission linewidth of the Lorentz-Bath susceptibility ($\gamma_0 + \gamma_{\rm{dephasing}}$) matched that of the Lorentz oscillator model ($\gamma$). The phenomenological relaxation rate of the bath oscillators was fixed to $\gamma_{\rm{b}}/2\pi = 10^{-2}\gamma/2\pi = 4\times 10^{-4}$.  A total of $10^2$ bath oscillators were included in each grid point described by the Lorentz-Bath susceptibility, with corresponding frequencies uniformly spanning the interval $[-5\gamma + \omega_0, 5\gamma + \omega_0]$. The specific implementation of the uniform or Lorentzian bath distribution was detailed in Sec. \ref{sec:implementation}.
    
    In the 1D polariton simulations, as shown in Fig. \ref{fig:demo}c, a planar Fabry--P\'erot cavity was formed by two dielectric mirrors separated by $d=1$ $\mu$m. Each mirror was represented by a  0.02 $\mu$m-thick dielectric layer with refractive index $n=10$. The aforementioned Lorentz or Lorentz-Bath dielectric slab was placed between the cavity mirrors to form strong coupling. Inside the cavity, the same broad-band Gaussian pulse was applied to excite the strong coupling system. During the simulations, the  EM energy confined between the mirrors was recorded, and the flux spectrum was computed to evaluate the polariton transmission.

    For the 2D polariton simulations in Fig. \ref{fig:2d_results}, the $y$-axis was explicitly included with a length of 20 $\mu$m, as depicted in Fig. \ref{fig:demo}c. The Bloch-periodic boundary conditions were imposed along the $y$-axis to reflect the translational invariance along this dimension. All other simulation details were kept identical to  those in the 1D case. For the MPI performance benchmark in Fig. \ref{fig:mpi_benchmark}, the spatial grid spacing was refined from $\Delta x=10^{-2}$ $\mu$m to $\Delta x=2.5\times10^{-3}$ $\mu$m, resulting in  $N_{\rm{grid}}=1.92\times 10^7$ spatial grid points --- an increase that facilitated efficient parallel execution across multiple compute nodes. 
    
    Note that for the multi-node MPI calculations, the default chunk dividing scheme provided by MEEP sometimes does not offer the optimal performance for the Lorentz-Bath model.
    To improve the multi-node performance in the Caviness HPC system hosted in the University of Delaware, the dynamic chunk balancing routine in MEEP was applied to balance the computational load in each node using real-time computational cost data across the nodes. In some other HPC architectures (such as the Anvil HPC system hosted in Purdue University \cite{Boerner2023}), the default chunk dividing scheme in MEEP may be sufficient.
    

%

    \end{document}